\documentclass{emulateapj}

\usepackage{amsmath,amsthm,amssymb,natbib}
\usepackage{graphicx}

\newcommand{\Nifs}{\ensuremath{^{56}{\rm Ni}}}

\newcommand{\kms}{\ensuremath{{\rm km~s^{-1}}}}

\newcommand{\Ms}{\ensuremath{M_{\rm s}}}
\newcommand{\vs}{\ensuremath{v_{\rm s}}}
\newcommand{\rs}{\ensuremath{r_{\rm s}}}
\newcommand{\rn}{\ensuremath{r_{\rm n}}}

\newcommand{\pb}{\ensuremath{p_{\rm b}}}
\newcommand{\ve}{\ensuremath{v_{\rm ej}}}

\newcommand{\vbo}{\ensuremath{v_{\rm bo}}}
\newcommand{\rbo}{\ensuremath{r_{\rm bo}}}
\newcommand{\tbo}{\ensuremath{t_{\rm bo}}}
\newcommand{\tr}{\ensuremath{t_{\rm tr}}}
\newcommand{\td}{\ensuremath{t_{\rm d}}}

\newcommand{\SNd}{SN-LSQ14BDQ}
\newcommand{\vsn}{\ensuremath{v_{\rm t}}}
\newcommand{\Msn}{\ensuremath{M_{\rm sn}}}
\newcommand{\Esn}{\ensuremath{E_{\rm sn}}}
\newcommand{\Lsn}{\ensuremath{L_{\rm sn}}}

\newcommand{\Ls}{\ensuremath{L_{\rm sh}}}
\newcommand{\Lsp}{\ensuremath{L_{\rm bo}}}

\newcommand{\Lm}{\ensuremath{L_{\rm m}}}
\newcommand{\Em}{\ensuremath{E_{\rm m}}}
\newcommand{\tm}{\ensuremath{t_{\rm m}}}
\newcommand{\tsn}{\ensuremath{t_{\rm sn}}}

\newcommand{\diff}[2]{\frac{d #1}{d #2}} 
\newcommand{\ergss}{\ensuremath{{\rm ergs~s^{-1}}}}

\newcommand{\zrho}{\ensuremath{\zeta_\rho}}
\newcommand{\zv}{\ensuremath{\zeta_{v}}}
\newcommand{\ztr}{\ensuremath{\zeta_{\rm tr}}}
\newcommand{\epss}{\ensuremath{\dot{\epsilon}_{\rm sh}}}

\newcommand{\Pms}{\ensuremath{P_{\rm ms}}}
\usepackage[]{natbib}

\received{}
\accepted{}

\shortauthors{Kasen, Metzger \& Bildsten}
\shorttitle{Magnetar Driven Shock Breakout}

\begin{document}
\bibliographystyle{apj}

\title{Magnetar Driven Shock Breakout and Double Peaked Supernova Light Curves}

\author{Daniel Kasen\altaffilmark{1,2}, Brian D. Metzger\altaffilmark{3}, Lars Bildsten\altaffilmark{4,5}}
\altaffiltext{1}{Nuclear Science Division, Lawrence Berkeley National Laboratory, Berkeley, CA, 94720, USA} 
\altaffiltext{2}{Departments of Physics and Astronomy, University of California, Berkeley, CA, 94720, USA}
\altaffiltext{3}{Columbia Astrophysics Laboratory, Columbia University, NY, NY, 10027, USA}
\altaffiltext{4}{Kavli Institute for Theoretical Physics, University of California, Santa Barbara, CA 93106, USA}
\altaffiltext{5}{Department of Physics, University of California, Santa Barbara, CA, 93106, USA}

\begin{abstract} 
The light curves of some luminous supernovae are suspected to be powered by the spindown energy of  a rapidly rotating magnetar. Here we describe a possible signature of the central engine: a burst of shock breakout emission occurring 
several~days after the supernova explosion.
The  energy input from the magnetar inflates a high-pressure bubble that drives a shock through the pre-exploded 
supernova ejecta.    If the magnetar is powerful enough, that shock will near 
the ejecta surface and become radiative. 
At the time of shock breakout, the ejecta will have expanded to a large radius ($\sim 10^{14}$~cm) so that the radiation released  is  at optical/ultraviolet wavelengths ($T_{\rm eff} \approx 20,000$~K) and lasts for several days. The luminosity and timescale of  this magnetar driven shock breakout are similar to the  first peak observed recently in the double-peaked light curve of \SNd.    However, for a large region of model parameter space, the breakout emission is predicted to be dimmer than the diffusive luminosity from direct magnetar heating.  A distinct double peaked light curve may therefore only be conspicuous
if   thermal heating from the magnetar is suppressed at early times.  We describe how such a delay in heating may naturally result from inefficient dissipation and thermalization of the  pulsar wind magnetic energy.  Without such suppression, the breakout may only be noticeable as a small bump or kink in the early luminosity or color evolution, or as a small but abrupt rise in the photospheric velocity.   A similar breakout signature may accompany other central engines in supernovae, such as a black hole accreting fallback material.
 \end{abstract} 
 
 \section{Introduction}
 
Optical surveys are finding a growing number of brilliant, though rare,
explosive transients, some $10-100$ times brighter than ordinary core collapse supernovae
\citep{Quimby_2007,Smith_2007, Ofek_2007,Barbary_2009,Pastorello_2010,Quimby_2011,Chomiuk_2011,Gal-Yam_2012,Howell_2013,Inserra_2013,Nicholl_2014, Papa_2015}.  The mechanism  generating  these enormous luminosities is unclear; the  energy sources that power ordinary supernova light curves -- the diffusion of shock deposited thermal
energy, or heating by radioactive \Nifs\ -- appear incapable of reproducing the observed
rise time and peak brightness of many of the  super-luminous supernovae (SLSNe).

Two classes of models are frequently invoked to explain the light curves of SLSNe.  The first involves interaction of the supernova ejecta with an extended circumstellar medium (CSM).  If interaction occurs  at a location where the ejecta is translucent (radii $\sim 10^{15}$~cm) the thermalized kinetic energy
can be radiated efficiently \citep{Woosley_2007,Smith_2007b,Chevalier_2011,Moriya_2011}.   In the second class of models, the ejecta from a seemingly ordinary supernova explosion
 is continuously reheated via energy injection from a long-lived central engine, either a rapidly rotating, highly magnetized
neutron star \citep[a millisecond magnetar;][]{Kasen_2010, Woosley_2010} or an accreting black hole \citep{Dexter_Kasen_2013}.  

Observations provide some evidence for both ideas.  In the  class of Type~II SLSNe, narrow ($\sim 10-100$~\kms) hydrogen Balmer lines
are often seen in emission, a signature of  interaction with a slow moving CSM.  In the Type~I SLSNe, on the other hand,
no spectroscopic indications of interaction are apparent and the line features only indicate rapidly moving
($\sim 10,000-15,000$~\kms) material.   
The magnetar powered model of SLSNe has been  successful in fitting the light curves, colors, photospheric velocity evolution, and gross
spectral features of several Type~I SLSNe \citep{Nicholl_2013, Inserra_2013, Dessart_2012, Howell_2013}.  
Late time observations of some Type~I SLSNe 
show emission continuing for 100's of days after the explosion, which has been claimed to be
indicative of persistent magnetar heating \citep{Inserra_2013}.

Additional observational tests are  needed to validate and  discriminate models of SLSNe. 
 Such an opportunity may  have arisen with  the well-sampled photometry of 
\SNd\ \citep{Nicholl_2015}, a Type~I SLSN with a double-peaked light curve.
The luminosity of \SNd\ rose to an  early maximum in $\approx 5$ days; then, after a brief  decline, the
light curve rose again
to an even brighter peak $(\approx 2 \times 10^{44}~\ergss$)  by $50$ days later.
 A similar double-peaked morphology had already been seen in the light curve of the SLSN SN2006oz \citep{Leloudas_2012}, although with poorer
 temporal sampling, and in SN~2005bf,
an unusual Type~Ib supernova of more ordinary brightness \citep{Anupama_2005,Folatelli_2006, Maeda_2007}.

Here we describe how a magnetar may produce a double-peaked  light curve.  The key insight is that 
a central engine heats the supernova ejecta in two physically and spatially
distinct ways. A thermalized pulsar wind heats the  ejecta directly  at its base, powering a luminosity that diffuses out on timescales of weeks or months.  At the same time, the pulsar wind dynamically affects the ejecta, 
 inflating a high-pressure bubble that drives shock heating at larger radii.  
If the magnetar is powerful enough, that shock will near 
the ejecta surface and become radiative, producing an early burst of emission.   

The situation resembles, in some ways, shock breakout from a stellar explosion \citep{Klein_1978, Matzner_Mckee_1999}, with a few key differences.  
First, a magnetar driven shock propagates through a moving medium; the shock will  be weaker, and when it does emerge, the  ejecta surface will have expanded by several orders of magnitude in radius.  The resulting emission will  last  longer and be at longer wavelengths (optical/UV) than  the brief x-ray burst that accompanies ordinary supernova shock breakout.
Second, the shock  does not necessarily die once it becomes radiative; as long as the magnetar continues to inject energy, the shell can be driven faster than free-expansion and may release energy at its outer edge.

The integrated  heating from a magnetar driven shock amounts to only a few percent of the total pulsar wind energetics.  However,   shock  heating occurs exterior to the bulk of the ejecta, and so can be radiated  $\lesssim 1$~week after explosion,  before most of the centrally thermalized  energy has had time to diffuse out.  Under certain circumstances, the two heating mechanisms  may produce two distinct emission maxima.
We develop below an analytic description of ``magnetar driven shock breakout", and present toy light curve calculations that suggest that this effect provides an appealing explanation for  double-peaked supernova light curves, and a   
means to constrain the magnetar model of SLSNe.

\section{Dynamics of Magnetar Driven Shocks}
  \label{sec:dyn}
  
Consider a pulsar with spin period $P$ and magnetic field $B$.  The total spin energy is
\begin{equation}
 \Em \approx 2\times 10^{52} \Pms^{-2}~{\rm ergs},
 \end{equation}
where $\Pms = P/1$~ms and we adopt a neutron star moment of inertia of $I = 10^{45}~{\rm g~cm^2}$. The rate at which energy is input from the pulsar is, in the case of vacuum magnetic dipole spindown
\begin{equation}
\Lm = \frac{\Em/\tm}{(1 + t/\tm)^2},
\label{eq:Lm}
\end{equation}
where the spindown timescale is
 \begin{equation}
 \tm \approx 5 B_{14}^{-2} \Pms^2~{\rm days},
 \end{equation}
where $B_{14} = B/10^{14}$~gauss and we assume, as in \cite{Kasen_2010}, that the angle between the rotation axis and magnetic dipole is $\alpha = 45^\circ$.   If the spindown energy is  thermalized in the ejecta, it may power a
supernova light curve \citep{Bodenheimer_1974,Gaffet_1977, Maeda_2007,Kasen_2010,Woosley_2010}.   The  dynamical effect of the energy injection, however, is independent of whether it thermalizes or not; in either case the energy behaves as a $\gamma = 4/3$ gas.

We will assume  that the supernova ejecta of mass \Msn\ is spherically symmetric and has a broken power law
density profile, with a shallow profile in the inner region and a steep one in the outer regions \citep{Chevalier_Soker_1989}. 
The  transition occurs at a velocity coordinate
\begin{equation}
v_t = \zv (\Esn/\Msn)^{1/2}.
\end{equation} 
 In the inner ejecta ($ v < \vsn$) the density at a position $r$ and time $t$ is
\begin{equation}
\rho(r,t) = \zrho \frac{\Msn}{\vsn^3 t^3} \left( \frac{r}{\vsn t} \right)^{-\delta}.
\label{eq:density}
\end{equation}
The density profile in the outer ejecta ($v > \vsn)$ has the same form but a different exponent, $\rho \propto r^{-n}$.    The coefficients are given by
\begin{equation}
\zrho = \frac{ (n-3)(3-\delta)}{4 \pi (n - \delta) } ,
\end{equation}
\begin{equation}
	\zv = \left[ \frac{2 (5-\delta)(n - 5)}{(n -3) (3 - \delta)} \right]^{1/2}.
\end{equation}
Typical values for core collapse supernovae are $\delta =1, n = 10$ \citep{Chevalier_Soker_1989}, which we adopt as  fiducial.

\begin{figure}
\includegraphics[width=3.5in]{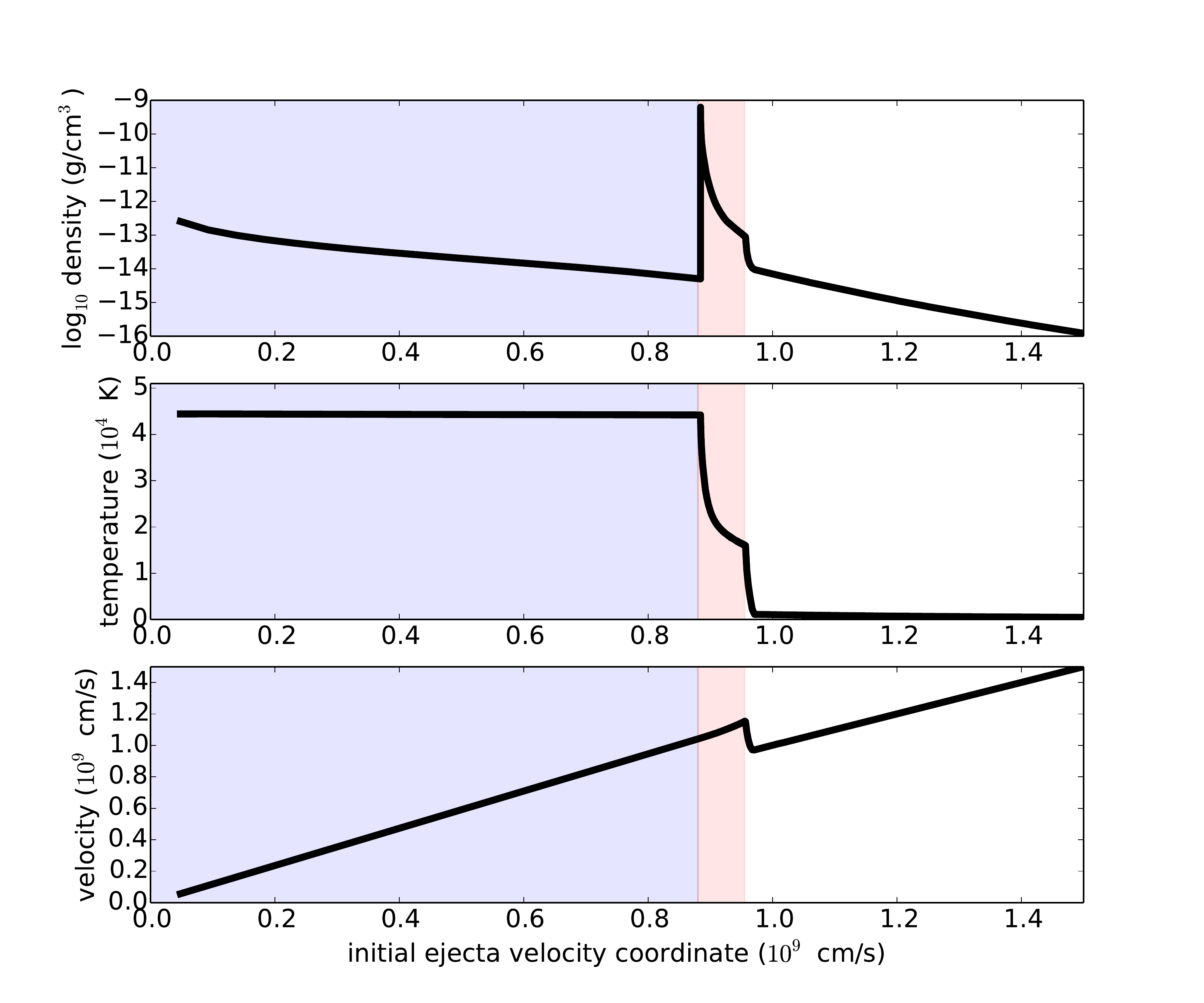}
\caption{1D hydrodynamical calculation of the dynamical effect of a magnetar on the supernova ejecta.  The ejecta had  $\Msn = 5~M_\odot$, and initially $\Esn = 10^{51}$~ergs and a broken power law density profile (Eq.~\ref{eq:density}). Energy injection from a magnetar with  $\Em = 5\times10^{51}$~ergs and $\tm = 5$~days was input according to Eq.~\ref{eq:Lm} in the inner few zones of the ejecta.  The snapshot shown is at a time $t = 15$~days.  The blue shading highlights the low density, high temperature magnetar bubble.  The red shading shows the region heated by the magnetar driven shock.   Radiation diffusion has not been included
in this calculation.
\label{fig:hydro}	}
\end{figure}

For $\Em \gtrsim \Esn$, the magnetar wind will  significantly restructure the supernova ejecta, as illustrated in
the hydrodynamical calculation of Figure~\ref{fig:hydro}.
The high pressure from central energy injection evacuates a cavity and
sweeps ejecta
into a thin shell. The shell moves  faster than 
the local ejecta expansion velocity, and a radiation dominated shock of relative velocity $\approx 2500~\kms$ forms.    
In a multi-dimensional calculation,  Rayleigh-Taylor instabilities would
break apart the shell and smear out the density peak \citep[e.g.,][]{Blondin_2001}, but the global structure would be qualitatively similar.
In addition, at late times, radiation diffusion from the inner hot bubble will smear out the temperature discontinuity at the shock front.

  \subsection{Time of Shock Emergence}

The magnetar driven shock  will become radiative when 
the diffusion time to the ejecta surface, $t_{\rm diff} \sim \tau R/c$, becomes comparable to the elapsed time,  $t \sim R/v$, or when the shock reaches an optical depth $\tau \sim c/v$.   
In the outer ejecta, the optical depth from the surface inward to velocity coordinate $v$ is
\begin{equation}
\tau(v) = \frac{1}{(n - 1)}\frac{c}{\vsn} \frac{\td^2}{t^2}  \left(\frac{v}{\vsn} \right)^{-n+1},
\label{eq:tau}
\end{equation}
where 
\begin{equation}
\td = \left[ \frac{ \zrho \Msn \kappa}{\vsn c} \right]^{1/2} \approx 32~M_{\rm sn,5}^{3/4} E_{\rm sn,51}^{-1/4} \kappa_{0.1}^{1/2}~{\rm days}.
\label{eq:td}
\end{equation}
is the  effective diffusion time  in a homologous expanding medium \citep{Arnett_1982}.   
Here $M_{\rm sn,5} = \Msn/5 M_\odot, E_{\rm sn,51} = \Esn/10^{51}$~ergs, and $\kappa_{0.1} = \kappa/0.1~{\rm cm^{2}~g^{-1}}$ is the scaled opacity. 
The shock then becomes radiative  ($\tau = c/v$) when it reaches a radius
\begin{equation}
\rbo = \vsn t \left[\frac{1}{\sqrt{n-1}} \frac{\td}{t} \right]^{2/(n-2)}.
\label{eq:r_bo}
\end{equation}
Eq.~\ref{eq:r_bo} assumes that $\rbo/t$ lies above the transition velocity coordinate \vsn, which is true for  $t < \td/\sqrt{n-1} \approx 10$~days.

To determine the time when the shock reaches the breakout radius \rbo, we make the assumption that mass is swept up into a geometrically thin shell \citep{Ostriker_Gunn_1971,Chevalier_1984, Chevalier_1992}.  The momentum and energy equations describing the shell dynamics  are
\begin{equation}
\Ms \diff{\vs}{t} = 4 \pi \rs^2[ \pb - \rho (\vs - \ve)^2],
\label{eq:dynP}
\end{equation}
\begin{equation}
\diff{(4 \pi \rs^3 \pb)}{t} = -4 \pi \pb r^2 \diff{\rs}{t} + \Lm(t) - L_{\rm sn}(t),
\label{eq:dynE}
\end{equation}
where \Ms, \rs, \vs\ are the mass, radius and velocity of the shell, $\ve = \rs/t$ is the ejecta velocity at radius \rs,  $\rho$ is the preshock ejecta density ahead of the shell, and \pb\ is the pressure in the magnetar inflated bubble.  The  \Lsn\ term represents the rate at which thermalized magnetar wind energy radiatively diffuses out of the bubble. 

Assuming that the magnetar injects energy at a nearly constant rate,  
$\Lm = \Em/\tm$ and that diffusion losses can be ignored ($\Lsn \approx 0$, appropriate for $t \ll \td$),  the dynamical equations have self-similar 
power law solutions for \rs\ and \pb\ \citep{Chevalier_1992}
\begin{equation}
r_s(t) 
= \vsn \tr^{1 - \alpha} t^{\alpha},
\label{eq:rsh}
\end{equation} 
where $\alpha = (6 - \delta)/(5- \delta)$ and 
\begin{equation}
\tr = \ztr \left(\frac{\Esn}{\Em} \right) \tm,
\end{equation}
is the  time it takes to the shell to propagate through the inner ejecta and reach the transition velocity \vsn.  
The coefficient is
\begin{equation}
\ztr = \left[ \frac{2 (n-5)(9 - 2 \delta)(11-2\delta)}{(5-\delta)^2 (n-\delta)(3-\delta)} \right]  .
\end{equation}
For $n=10,\delta = 1$, we have $\alpha = 5/4$ and $\ztr = 2.2$.  
The expression for \tr\ has been previously derived in
\cite{Chevalier_2005}.

This self-similar power law solution only  holds for times $t < \tr$  when the shell remains in the inner ejecta.  At later times, the shell moves into the steep outer ejecta and  accelerates.     The shock front will begin to move ahead of the shell,  but the thin shell approximation appears to remain reasonably   valid (see Figure~\ref{fig:hydro}).  The limiting behavior can be understood by noting that,
at large radius, nearly all of the mass is swept into the shell, $\Ms \rightarrow \Msn$.
Given the low densities in the outer ejecta,  the ram pressure term  $\rho (\vs - \ve)^2$ can  be neglected relative to the bubble pressure, and  the dynamical equations give asymptotically a power law (Eq.~\ref{eq:rsh}) with
  $\alpha = 3/2$.

\begin{figure}
\includegraphics[width=3.5in]{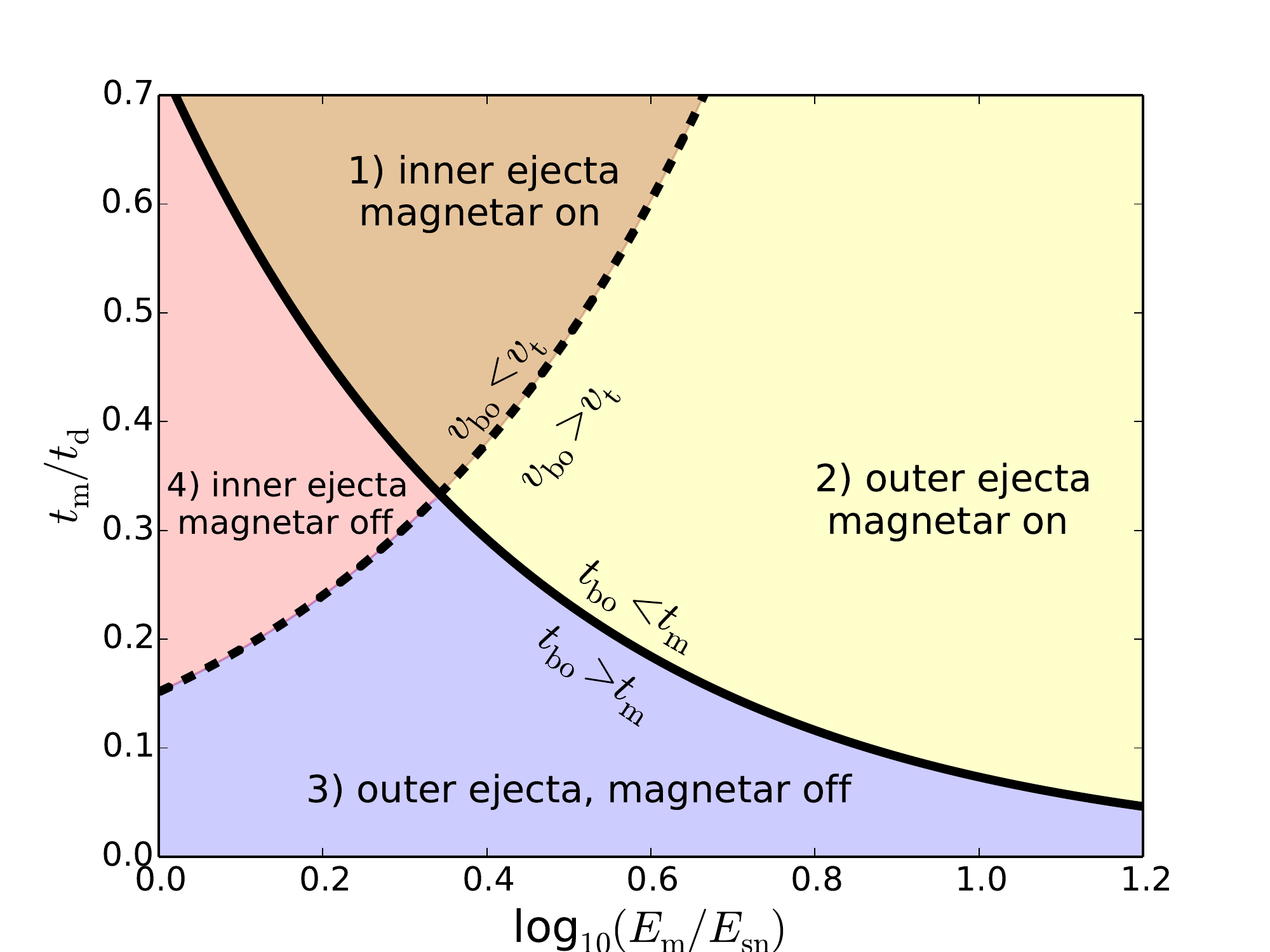}
\caption{The different regimes of magnetar driven shock emergence, illustrating whether the  shock becomes radiative in the inner ejecta ($\vbo < v_t$) or outer ejecta ($\vbo > v_t$) 
and whether the magnetar is ``on" ($\tbo < \tm$) or ``off " ($\tbo > \tm$) at the time of emergence. The solid lines are
determined from Eqs.~\ref{eq:tr_tsh} and \ref{eq:tm_tsh} with $\alpha = 5/4$ and $n = 10$.  Breakout emission can occur in all regions, but will be brightest in region 2, when the shock is strongest.
  \label{fig:regions}}
\end{figure}

Although the break in the density profile complicates the dynamics, the
shell radius can be reasonably approximated by $\rs \propto t^\alpha$ where the exponent is in the range $\alpha \approx  1.2 - 1.5$.  If we assume that the
shock becomes radiative in (or near) the outer ejecta,  the time of shock breakout is found by setting
 $\rs$ equal to the 
breakout radius, \rbo\ (Eq.~\ref{eq:r_bo}), giving 
\begin{equation}
\tbo \approx \left[ \frac{ \ztr^{1-\beta}}{(n-1)^{\beta/2}}\right] \td^\beta \tm^{1- \beta} (\Esn/\Em)^{1 - \beta},
\label{eq:tbo}
\end{equation}
where
\begin{equation}
\beta = \frac{2}{(\alpha - 1)(n-2) + 2}.
 \end{equation}
 The shock breakout time, \tbo, is a weighted geometric mean of \td\ and \tm, where
the appropriate value of $\beta$ depends on whether the shell has spent most of its time in the inner or outer regions of ejecta, i.e., whether \tbo\ is much greater than or less than \tr. 
For $\tbo \lesssim\tr$, the self-similar value $\alpha \approx 5/4$ applies, and $ \beta = 1/2$ (for $n=10$).
For  $\tbo \gg \tr$, the asympotic value $\alpha = 3/2$ is 
more accurate and $\beta = 1/3$.
The distinguishing condition is
\begin{equation}
\tbo \gtrsim \tr~~{\rm if}~~\Em \gtrsim 6.6 \Esn  \frac{\tm}{\td}.
\label{eq:tr_tsh}
\end{equation}

To further complicate  the dynamics, at times $t > \tm$, the energy input from the magnetar ``shuts off", i.e., \Lm\ drops towards zero.  Following this, the bubble pressure  decreases nearly adiabatically and the shell radius approaches free expansion, $\rs \propto  t$.  A shock breakout  can still occur, but the emission will be less luminous because the shock  speed declines for $t > \tm$.   The brightest shock breakouts then occur when the shock  becomes radiative before the magnetar shuts off, which requires a magnetar energy
\begin{equation}
\tbo < \tm~~{\rm if}~~\Em \gtrsim  0.7 \Esn 
\left( \frac{\td}{\tm} \right)^{\beta/(1 - \beta)}.
\label{eq:tm_tsh}
\end{equation}

The two conditions (Eqs.~\ref{eq:tr_tsh} and \ref{eq:tm_tsh}) divide the \Em-\tm\ parameter space  of shock emergence
into four regimes,  illustrated in Figure~\ref{fig:regions}.  
The partitioning is only suggestive, as  real ejecta density profiles are
more complex than a broken power law, and the magnetar
energy deposition does not shut off sharply, but rather follows a smooth function of the form Eq.~\ref{eq:Lm}.
The most luminous breakouts occur when the shock is being continuously driven through the steep outer ejecta (region~2), which happens when the magnetar is energetic and/or long lasting.  On the other extreme, for $\Em \ll \Esn$ or $\tm \ll \td$ the shock will stall out before being revealed and no prominent shock breakout signature is expected..

\begin{figure}
\includegraphics[width=3.5in]{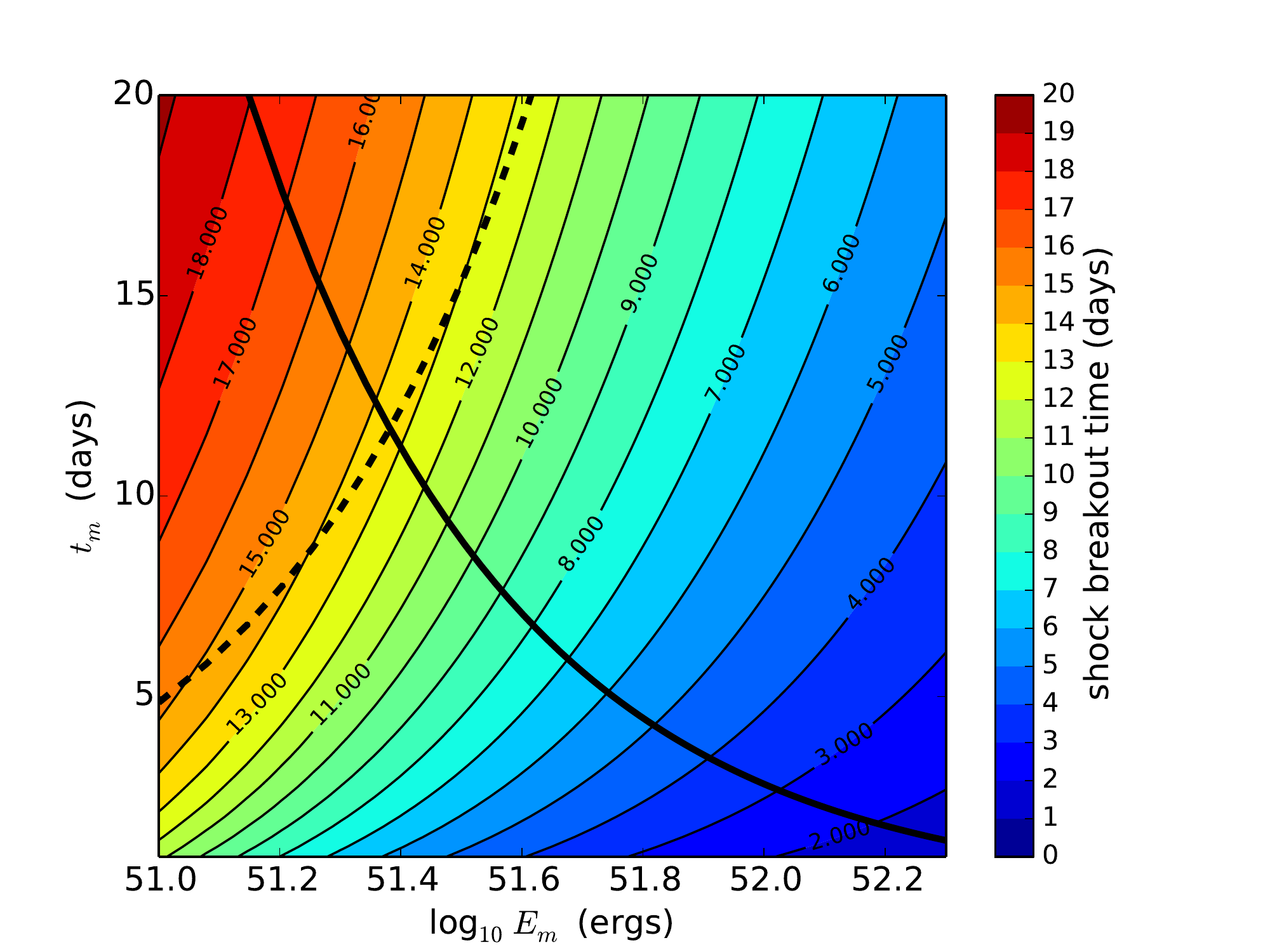}
\caption{Numerical  calculation  of the shock breakout time, \tbo, for various values of \Em\ and  \tm, and standard ejecta parameters: $\Msn = 5~M_\odot$, $\Esn = 10^{51}$~ergs,  $\delta = 1, n = 10$, $\kappa = 0.1$~g~cm$^2$, 
which imply a diffusion time $\td = 32$~days. The magnetar energy
injection rate was taken from Eq.~\ref{eq:Lm}.  The solid line shows the estimated $\tbo = \tm$ contour from Eq.~\ref{eq:tm_tsh}; the dashed line shows the estimated $\vbo = \vsn$ contour from Eq.~\ref{eq:tr_tsh}.
\label{fig:tsh}}
\end{figure}

To better determine the time of shock emergence, 
we numerically integrated the dynamical equations (Eqs.~\ref{eq:dynP} and \ref{eq:dynE}) to follow the shell location
 and find when it reached the $\tau = c/v$ surface.  This calculation did not assume $\Lsn = 0$, but rather included approximate radiative loses
via the method described in the Appendix. 
 Figure~\ref{fig:tsh} shows this calculation of \tbo\ for various values of \Em\ and \tm, and
  standard ejecta parameters.
When we carry out the numerical integration using \Lm\  constant for $t < \tm$, and zero afterwards, the resulting \tbo\
agrees well with the  analytic expression Eq.~\ref{eq:tbo}.  For the more realistic case of continuous energy injection (given by Eq.~\ref{eq:Lm}) the analytic result underestimates \tbo\ by about 20\%.  In this case, a better  estimate is achieved by multiplying  \tm\ in Eq.~\ref{eq:tbo} by a factor $\approx 1.5$ to account for  the non-zero magnetar energy injection at later times.

\subsection{Shock Heating}
\label{sec:shock}

We next estimate the local heating from the magnetar driven shock, which will set 
the luminosity of breakout when the shock emerges.  The rate at which energy is dissipated at the shock is 
\begin{equation}
\epss(t) = 4 \pi \rs^2\frac{ \rho}{2} (\vs - \ve)^3  = 4 \pi \rs^2   \frac{ \rho}{2} \ve^3 \eta^3,
\label{eq:esh}
\end{equation}
where in the second expression we have parameterized the  shock velocity as
\begin{equation}
\eta(t) = \frac{\vs - \ve}{\ve} = 	\frac{t}{\rs} \left( \diff{\rs}{t} - \frac{\rs}{t} \right) .
\label{eq:eta}
\end{equation}
The shock strength  parameter $\eta$ will play an important role in determining the radiated luminosity.  
If the shell radius obeys a powerlaw, $\rs \propto t^\alpha$ then $\eta = \alpha - 1$.
In the inner ejecta, $\alpha$ approaches the self-similar value
$\alpha = (6 - \delta)/(5 - \delta)$ and
  $\eta = 1/(5 -\delta)$, or $\eta = 1/4$ for $\delta =1$.    In other words, the  shell moves $25\%$ faster than 
  the local ejecta velocity.
  For $t \gg \tr$, when the shock propagates deep into  the steep outer ejecta and accelerates, $\alpha \rightarrow 3/2$ and 
  $\eta \rightarrow 1/2$.

\begin{figure}
\includegraphics[width=3.7in]{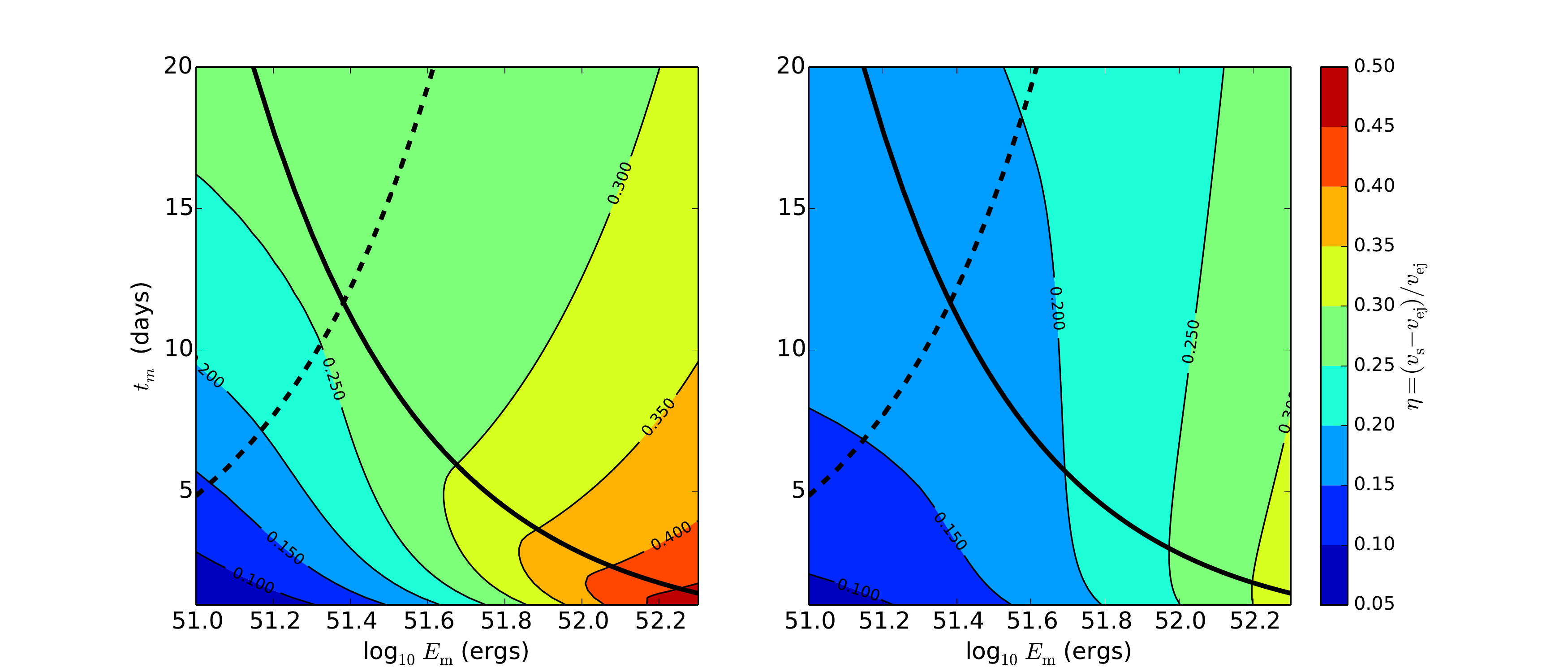}
\caption{Numerical  calculation of the shock strength parameter $\eta = (\vs - \ve)/\ve$ at the time of shock breakout for various values of  \Em\ and \tm, and  standard ejecta parameters: $\Msn = 5~M_\odot$, $\Esn = 10^{51}$~ergs,  $\delta = 1, n = 10$, $\kappa = 0.1$~cm$^2$~g$^{-1}$. The left panel assumes a constant magnetar energy injection rate $\Lm = \Em/\tm$ for $t < \tm$ and $\Lm = 0$ afterwards.  The right panel uses the continuous 
injection rate of Eq.~\ref{eq:Lm}.  
The solid line shows the estimated $\tbo = \tm$ contour from Eq.~\ref{eq:tm_tsh}; the dashed line shows the estimated $\vbo = \vsn$ contour from Eq.~\ref{eq:tr_tsh}.
\label{fig:eta}}
\end{figure}

Using our numerical integration of the dynamical equations discussed in Section~\ref{sec:dyn}, we  calculated the value of $\eta$ at
the time of  shock emergence.
The left side of Figure~\ref{fig:eta} shows the numerical determination of $\eta$ for the simple case where
$\Lm = \Em/\tm$ is constant for  $t < \tm$, then immediately drops to $\Lm = 0$.  The behavior follows  analytical expectations: in  region 1 (shock emergence in the inner ejecta, magnetar on)  $\eta = 0.25$; in region 2 (shock emergence  in the outer ejecta, magnetar on) $\eta > 0.25$ and increases with increasing magnetar energy, approaching a maximum value $\eta = 0.5$.  For region 3 (shock emergence in the outer ejecta, magnetar off) $\eta$ declines as $\tm$ decreases, illustrating the progressive weakening of the shock following magnetar shut off. 

The right panel of Figure~\ref{fig:eta} shows the behavior of $\eta$ in a calculation using continuous magnetar energy injection  given by
Eq.~\ref{eq:Lm}. Similar trends with \Em\ and \tm\ are seen, but the values of $\eta$ are generally lower, as 
the magnetar energy deposition is spread out over a longer timescale, with \Lm\ always less than
$\Em/\tm$.

\begin{figure}
\includegraphics[width=3.0in]{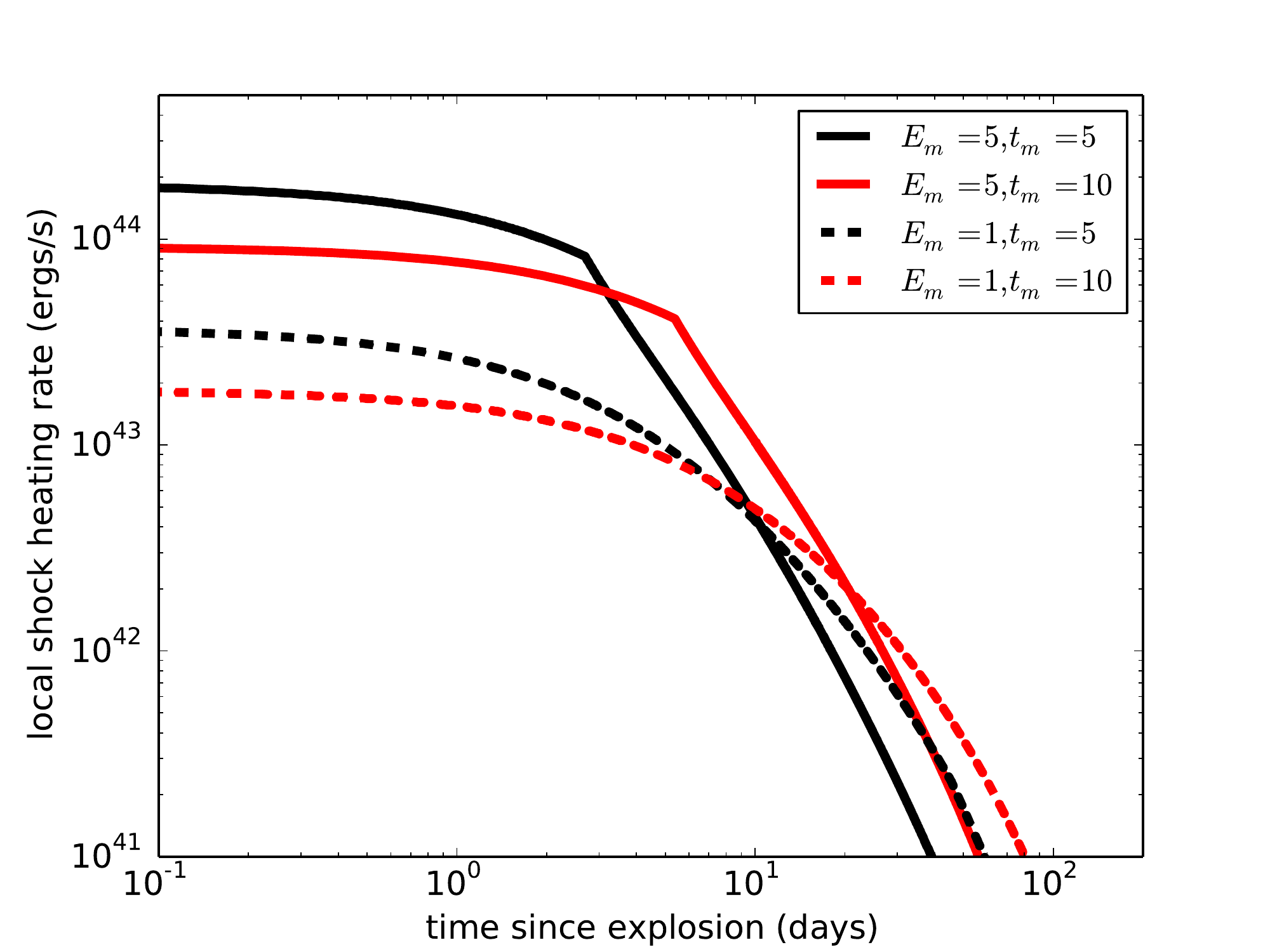}
\caption{Calculation of  the local shock heating rate (Eq.~\ref{eq:esh}) over time for standard ejecta parameters $\Msn = 5~M_\odot$, $\Esn = 10^{51}$~ergs,  $\delta = 1, n = 10$, and continuous magnetar energy injection  given by
Eq.~\ref{eq:Lm}. The legend
gives the value of the magnetar energy (in units of $10^{51}$~ergs) and the spindown time (in days).  A power-law break in the heating rate is seen when the shock enters the steep density profile of the outer ejecta.  The luminosity of shock breakout is approximately the local heating rate at the time of  breakout.
\label{fig:shock}}
\end{figure}

Figure~\ref{fig:shock} shows the time dependence of the local shock heating rate (Eq.~\ref{eq:esh})
for continuous magnetar energy injection.   The luminosity of the shock breakout pulse will 
 approximately equal the local heating rate at the time of  breakout.
 The local heating falls off  with time roughly like a power-law, due to the 
progressive decrease in the preshock density.  A break in the local heating rate is seen when the shock enters the steep density profile of the outer layers ($v > \vsn$).  
The shock is weakened at times $t > \tm$ by the decrease in magnetar energy injection, and further at times $t \gtrsim \td$ by radiation diffusion through the ejecta shell, which  depressurizes the magnetar bubble.  This latter effect is included in the calculation by including a non-zero diffusion term \Lsn\ in Eq.~\ref{eq:dynE}.  

\section{Observational Consequences}

\subsection{Properties of Shock Breakout}

To estimate the peak luminosity resulting from shock heating, we can use  ``Arnett's law" \citep{Arnett_1982} which states that, for any general heating source, the luminosity at the light curve peak is equal to the instantaneous rate of energy deposition at that time.
We thus determine the peak luminosity by evaluating the heating rate Eq.~\ref{eq:esh} at the shock breakout time, \tbo
\begin{equation}
\Lsp = 2 \pi \zrho \frac{\Msn \vsn^2}{\tbo}  \eta^3 \left( \frac{\vbo}{\vsn} \right)^{-n + 5}.
\label{eq:Lso}
\end{equation}
Shock breakout occurs at a velocity coordinate $\vbo = \rbo/\tbo$ determined from Eqs.~\ref{eq:r_bo} and \ref{eq:tbo} 
\begin{equation}
\vbo \approx  1.1 \vsn \left[ \frac{\td}{\tm} \frac{\Em}{\Esn} \right]^{\frac{2(1-\beta)}{n-2}}.
\end{equation}
Evaluating Eq.~\ref{eq:Lso} at the velocity coordinate \vbo\ then gives the peak luminosity
\begin{equation}
\Lsp \approx 9 \frac{\Esn }{\td}  \eta^3 \left[ \frac{\tm}{\td} \frac{\Esn}{\Em} \right]^{\frac{(n-8)(1-\beta)}{(n-2)}} .
\label{eq:Ls2}
\end{equation}
For standard parameters ($\beta = 1/2, n=10$) the exponent of the term in brackets is only $1/8$, and we see that the peak luminosity depends most sensitively on the shock strength parameter, $\eta$.  The luminosity also depends on \Esn, as this sets the ejecta expansion velocity and hence the size of the remnant at the time of breakout.

The spectrum of the breakout emission can be approximated by a quasi-blackbody
with an  effective temperature, $T_{\rm eff}$, 
determined by $\Lsp= 4 \pi r_{\rm p}^2 \sigma T_{\rm eff}^4$, where
$r_{\rm p}$ is the photospheric radius defined by the $\tau = 1$ surface. From  Eq.~\ref{eq:tau}
\begin{equation}
r_{\rm p} \approx 1.2 \vsn \td (\tbo/\td)^{(n-3)/(n-1)}. 
\label{eq:rp}
\end{equation}

Plugging  fiducial parameters ($n = 10, \delta = 1, \beta = 1/2$) into Eqs.~\ref{eq:tbo}, \ref{eq:Ls2}, and \ref{eq:rp}  we arrive at analytic estimates of the  time, luminosity, photospheric radius, and effective temperature at the time of a magnetar driven shock breakout
\begin{eqnarray}
\tbo &\approx& 10.8 ~M_{\rm sn,5}^{3/8} E_{\rm sn,51}^{1/4}  E_{\rm m,51}^{-1/2} t_{\rm m,5}^{1/2}~{\rm days} \\
\Ls &\approx& 2.1\times10^{43} ~\eta_{0.2}^3  M_{\rm sn,5}^{-1/2} E_{\rm sn,51}^{3/4} E_{\rm m,51}^{-1/8} t_{\rm m,5}^{1/8}~\ergss \\
r_{\rm p} &\approx& 7.6 \times 10^{14}  M_{\rm sn,5}^{-0.04} E_{\rm sn,51}^{0.64} E_{\rm m,51}^{-0.39} t_{\rm m,5}^{0.39}
\kappa_{0.1}^{0.11}~{\rm cm} \\
T_{\rm eff} &\approx& 15,500~ 
\eta_{0.2}^{0.75} M_{\rm sn,5}^{-0.1} E_{\rm sn,51}^{-0.13} E_{\rm m,51}^{0.16} t_{\rm m,5}^{-0.16} \kappa_{0.1}^{-0.06}~{\rm K}.
\end{eqnarray}
where $E_{\rm sn,51} = \Em/10^{51}$~ergs, $t_{\rm m,5} = \tm/5$~days, $\eta_{0.2} = \eta/0.2$.
In most stellar shock breakout events, the duration of the burst is set by the light crossing time of the ejecta, however in the present case the ejecta is extended and the timescale is set by the diffusion time, \tbo\ \citep[see e.g.,][]{Chevalier_2011, Piro_2013}.

\begin{figure}
\includegraphics[width=3.5in]{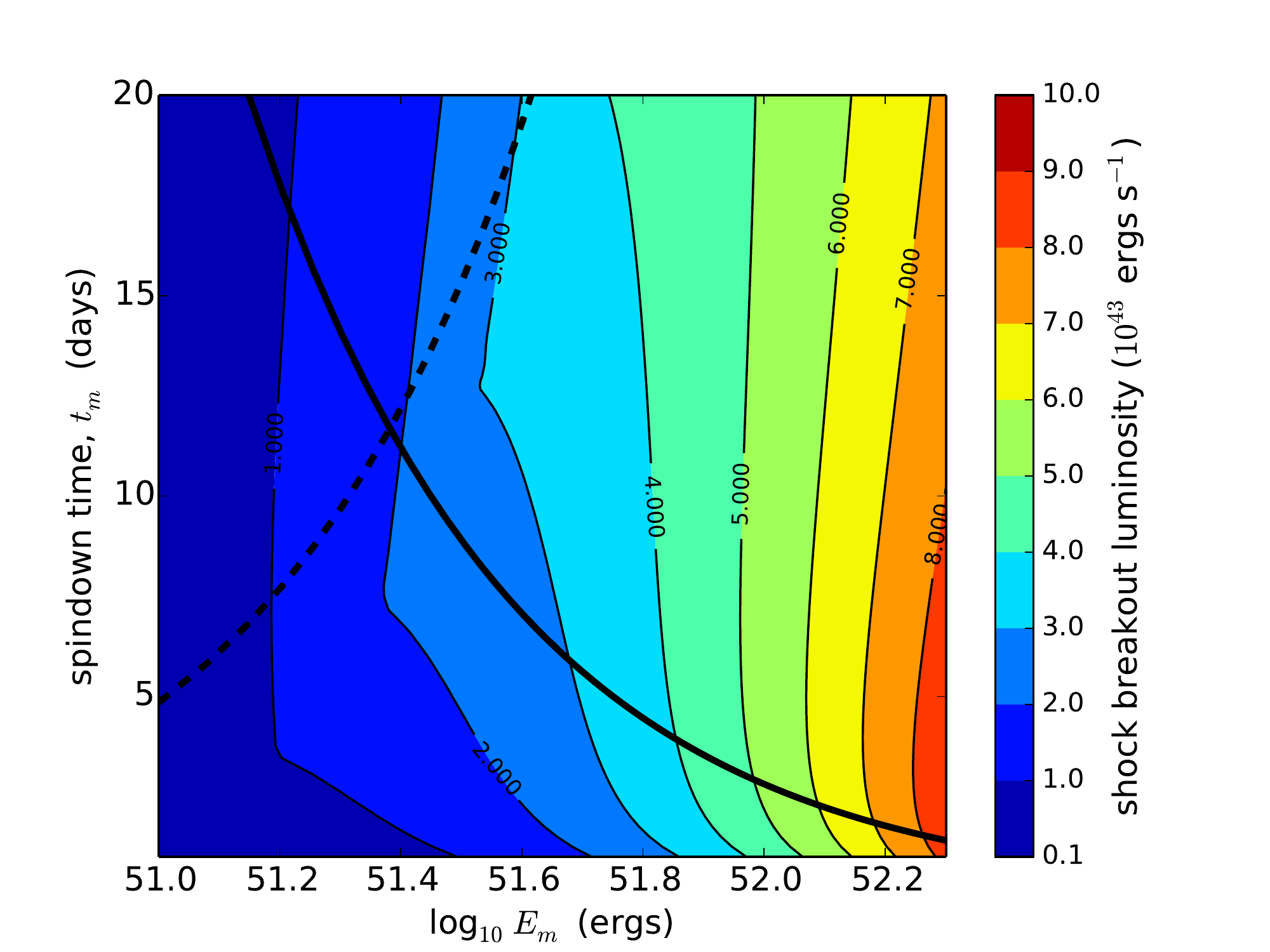}
\caption{Same as Figure~\ref{fig:tsh} but showing the calculated luminosity at the time of shock breakout.
\label{fig:Lsh}
}
\end{figure}

The analytic results are only approximate; as an improved estimate of the peak luminosity, we used our numerical integration  of the shell evolution to evaluate the heating rate at the time of  shock emergence.  Figure~\ref{fig:Lsh} shows
the numerical results in the \Em-\tm\ parameter space.  The luminosity increases with \Em\ as higher
magnetar energy drives a stronger shock (greater $\eta$). For low values of \tm, the
the luminosity drops due to the decline of $\eta$ following magnetar shut off.
\\
\\

 \subsection{Approximate Light Curves}
 The light curve of a magnetar powered supernova will be the sum of the emission from  shock breakout and the
diffusive luminosity from central magnetar heating.    A  first peak will only be distinguishable  when the shock breakout  luminosity  is comparable to or greater than the diffusive  luminosity at that time.  To model the composite light curve, we  used a one-zone formalism \citep{Arnett_1982} to calculate approximate emission from each mechanism, then  added the results.  The method is described in the Appendix.

The left panel of Figure~\ref{fig:LC} shows a composite model bolometric light curve.  For reasonable magnetar and ejecta parameters ($\Em = 5\times10^{51}~{\rm ergs}, \tm = 2~{\rm days}, \Msn = 5~M_\odot$, $\Esn = 10^{51}$~ergs), 
the shock breakout emission is not dominant and  produces only a kink in the early light curve rise.   The lack of a prominent first peak is consistent with the grey radiation-hydrodynamical calculations of \cite{Kasen_2010}.  Despite the lack of a distinct light curve bump, the breakout may still be detectable by the shift in brightness and colors at the time of breakout, or by a sudden, but small ($\sim 1000$~\kms) increase in the line velocities when the photosphere recedes through the region of non-monotonic velocity.

To see a clear double-peaked light curve requires either a very bright shock breakout, or a slowly evolving diffusive light curve,
properties that are only realized in certain regions of parameter space.
Figure~\ref{fig:LC_range} shows that increasing the ejecta mass  delays the diffusive light curve, making the shock breakout peak more prominent.   Increasing the  kinetic energy of the supernova explosion leads to a larger remnant and brighter shock emission, which also clarifies the double-peaked structure.  In addition, for very low supernova and magnetar energies, $\Esn \approx \Em \approx 10^{50}$~ergs, the main light  curve evolves slowly and a low luminosity double-peaked light curve can be seen.

Three additional physical effects may further distinguish the shock breakout peak: 1) Though the model light curves  here and in \cite{Kasen_2010} assume a constant grey opacity,  the true ejecta opacity likely increases inwards, given the higher temperature and ionization state of the inner regions, and the possible presence of synthesized iron group elements.    A higher opacity in the  interior would delay the diffusive light curve relative to the shock breakout emission.  2) Deviations from spherical symmetry due to bipolar magnetar energy injection may lead to breakout occurring first along the poles, making the breakout emission more conspicuous from polar viewing angles.  3) Inefficient thermalization of the magnetar wind would delay the rise of the diffusive light curve; we discuss this  point in more detail in the next section.

\begin{figure}
\includegraphics[width=3.5in]{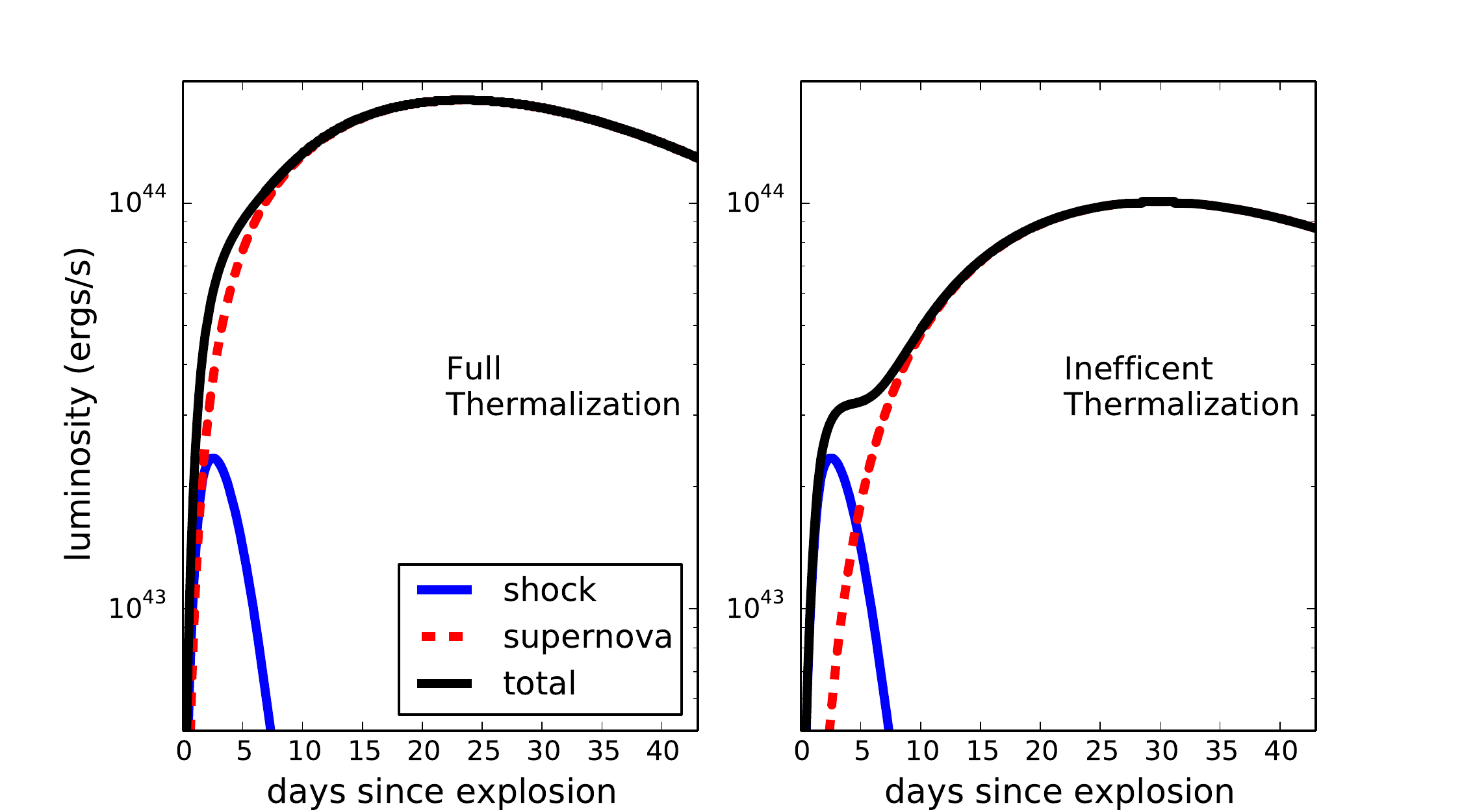}
\caption{Approximate model light curves  including both shock breakout emission (blue lines) and diffusive luminosity from magnetar heating (red dashed lines).  The magnetar parameters are
$\Em = 5\times 10^{51}$~ergs and $\tm = 2$~days, corresponding to $B_{14} = 4, P = 2.5~{\rm ms}$.  The supernova ejecta had kinetic energy
 $\Esn = 1 \times 10^{51}~{\rm ergs}$,  mass 5~$M_\odot$, and opacity $\kappa = 0.1~{\rm cm^2~g^{-1}}$.  The right panel
 assumes 100\% thermalization of magnetar energy; the left panel assumes inefficient thermalization according to Eq.~\ref{eq:Lth} with $Y = 0.1, A = 0.9$.
   \label{fig:LC} }
\end{figure}

\begin{figure}
\includegraphics[width=3.5in]{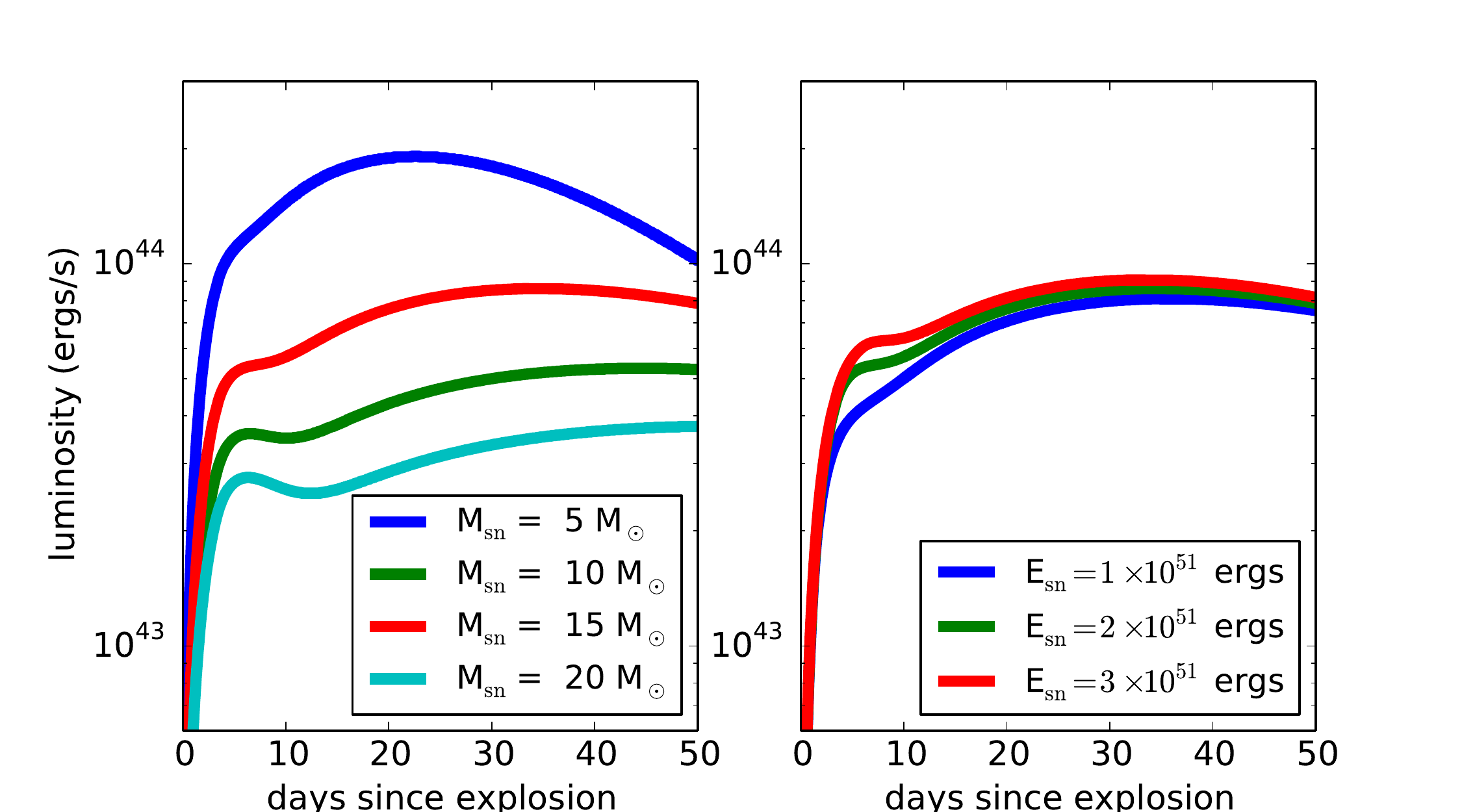}
\caption{Approximate model light curves  including both shock breakout emission and diffusive luminosity from magnetar heating.   The magnetar parameters are
$\Em = 5\times 10^{51}$~ergs and $\tm = 2$~days, corresponding to $B_{14} = 4, P = 2.5~{\rm ms}$.  In the left panel, the supernova kinetic energy is $\Esn = 10^{51}$~ergs and the mass is varied from $\Msn = 5 - 20~ M_\odot$.  In the right panel, the supernova mass is $\Msn = 10 M_\odot$
and the kinetic energy is varied from $\Esn = 1-3 
\times 10^{51}~{\rm ergs}$.  The calculations assume 100\% efficient magnetar thermalization and opacity $\kappa = 0.1~{\rm cm^2~g^{-1}}$.
   \label{fig:LC_range} }
\end{figure}

\section{Magnetar Wind Thermalization}
\label{sec:thermalize}

If the magnetar wind does not thermalize efficiently at early times, 
the rise of the diffusive light curve will be delayed, which will clarify the shock-breakout peak.  There are physical reasons to think this delay in thermalization may occur in SLSNe.

The spin-down luminosity of a magnetar is initially carried outwards by
a Poynting flux.  The magnetic field is  initially strong close to
magnetar surface, becoming dominated by its toroidal component outside of
the light cylinder radius.   Eventually, reconnection inside the nebula
(e.g.~\citealt{Porth+13,Mosta+14,Bromberg15}) will covert
the magnetic energy to high energy  $e^{\pm}$ particles.  However, prior to the (uncertain) timescale for reconnection,  the magnetar wind energy is not inefficiently thermalized, and takes the form of a magnetically dominated outflow that drives a shock through the ejecta.  

Even after reconnection dissipates the magnetic field energy, other physical
effects may reduce the thermalization efficiency for a continued period of time.
The dissipation of the wind energy at the termination shock or reconnection layers
generates primarily high energy
$e^{\pm}$ pairs.  The  injected pairs cool rapidly via synchrotron and inverse Compton radiation, producing  high energy ($\gg m_e c^{2}$) photons that may in  turn generate additional $e^{\pm}$ pairs by interacting with  background thermal radiation  \citep{Metzger+14}.  The optical depth to $\gamma-\gamma$ interactions is quantified by the compactness parameter,
\begin{eqnarray}
\ell &\equiv& \frac{E_{\rm nth}\sigma_{\rm T} \rn}{V_{\rm n}m_e c^{2}} \approx 2000 L_{\rm m,45}t_{\rm day}^{-2}v_{9}^{-2}
\label{eq:compactness}
\end{eqnarray}
where $L_{\rm,m,45} = L_{\rm m}/10^{45}~\ergss$, $v_9 = v_{\rm t}/10^9~{\rm cm~s^{-1}}$, $t_{\rm day} = t/1~{\rm day}$, $\sigma_{\rm T}$ is the Thomson cross-section, $E_{\rm nth} \sim L_{\rm m}t$ is the approximate non-thermal energy injected into the nebula by the magnetar on the expansion time, $V_{\rm n} \approx 4\pi \rn^{3}/3$ is the volume of the nebula, and $\rn \sim \vsn t$ is the approximate nebula radius. 

If $\ell \gg 1$ then pairs produced by the first generation of photons upscatter additional seed photons to sufficient energies to create additional pairs.  The details of this `pair cascade' are complex, but the net effect is to convert a sizable fraction $Y$ of the injected spin-down power into $e^{\pm}$ pairs (e.g., \citealt{Svensson87}).   The total number of pairs $N_{\pm}$ in the nebula is set by the equilibrium between the rate of pair creation and annihilation
\begin{equation}
\dot{N}_{\pm}^{\rm +} \simeq \frac{Y L_{\rm m}}{m_e c^{2}} ~~~{\rm and}~~
\dot{N}_{\pm}^{\rm -} = \frac{3}{16}\sigma_{\rm T} c N_{\pm}^{2}V_{\rm n}^{-1}.
\label{eq:Ndotminus}
\end{equation}
In equilibrium ($\dot{N}_{\pm}^{\rm +} = \dot{N}_{\pm}^{\rm -}$), the Thomson optical depth of pairs across the nebula is \begin{eqnarray}
\tau_{\rm es}^{\rm n} &=& \sigma_{\rm T} \rn n_{\pm} = 
 \left[\frac{4 Y\sigma_{\rm T} L_{\rm m}}{\pi \rn m_e c^{3}}\right]^{1/2} \approx 19 Y^{1/2}L_{\rm m,45}^{1/2}v_{9}^{-1/2}t_{\rm day}^{-1/2}
\label{eq:taueseq}
\end{eqnarray}
This equilibrium is reached on a timescale, $t_{\rm eq} \simeq 16 \rn/3 c \tau_{\rm es}^{\rm n}$, which is short compared to the evolution timescale as long as $\tau_{\rm es}^{\rm n} \gg 16 \vsn /3c$.  

  Most of the remaining fraction $1-Y$ of the energy released by the cooling $e^{\pm}$ pairs goes into a non-thermal power-law tail of radiation. 
  The high  scattering optical depth $\tau_{\rm es}^{\rm n}$ of the nebula  traps these  photons and delays their thermalization.  On average, a hard photon must interact with the nebula walls $\sim (1-A)^{-1}$ times before thermalizing, where the albedo $A$ is the probability that the photon is scattered back into the nebula instead of being absorbed by the walls.  The `lifetime' of a hard photon is therefore  $t_{\rm life} = t_{\rm d}^{\rm n}(1-A)^{-1}$, 
where $t_{\rm d}^{\rm n} \simeq (\tau_{\rm es}^{\rm n}+1) \rn/c$
is the photon diffusion time required for a single nebula crossing.  

If $t_{\rm life}$ exceeds the expansion time, then non-thermal photons lose energy to adiabatic expansion before their energy can be thermalized. 
 This reduces the effective rate of thermal energy production to a fraction of the magnetar spin-down power (\citealt{Metzger&Piro14})
\begin{equation}
L_{\rm th} = \frac{L_{\rm m}}{1 + (t_{\rm life}/t)},
\label{eq:Lth}
\end{equation}
where
\begin{equation}
\frac{t_{\rm life}}{t} = \frac{\tau_{\rm es}^{\rm n}v_{\rm t}}{c(1-A)} \approx 0.6\frac{Y^{1/2}}{1-A}L_{\rm m,45}^{1/2}v_{9}^{1/2}t_{\rm day}^{-1/2}
\end{equation}
and we have assumed $\tau_{\rm es}^{\rm n} \gg 1$.

Depending on the characteristic values of $Y$ and $A$, suppression of the magnetar power due to thermalization can be important.  A typical value of the pair multiplicity is $Y \sim 0.1$ (\citealt{Svensson87}), although its precise value will depend on the nature of the pair cascade and deserves further study.  The albedo depends on the ionization parameter, which
sets the ratio of scattering to absorption in the ejecta wall.
Photoionization calculations by \citet{Ross+99} (their Fig.~2) show a rather high albedo $A \sim 0.9$ across a range of photon energies $\sim 1-30$ keV, although for high photon energies inelastic scattering results in a higher absorbed fraction.\footnote{Also note that we have assumed that photons absorbed by the ejecta are immediately thermalized.  However, in reality if their energy is instead deposited in the electrons in the hot outer ionized layer (where the Compton temperature is much higher), additional Compton down scattering may be required to diffuse this energy to optical/UV wavelengths, resulting in an effective value of $A$ which is even higher.}

For values $Y = 0.1, A = 0.9$,  thermalization of the magnetar wind will be suppressed for  $\sim 1$~week 
following the explosion.   Inefficient thermalization will not affect the dynamics of the shock, but will reduce the
early diffusive luminosity from direct magnetar heating.  We included this effect in our light curve calculations by using the suppressed 
magnetar luminosity Eq.~\ref{eq:Lth} to determine the supernova light curve.  The results, shown in the right panel of
Figure~\ref{fig:LC} demonstrate that inefficient thermalization may serve to better distinguish the
double-peaked light curve shape.

\section{Discussion and Conclusions}

\begin{figure}
\includegraphics[width=3.5in]{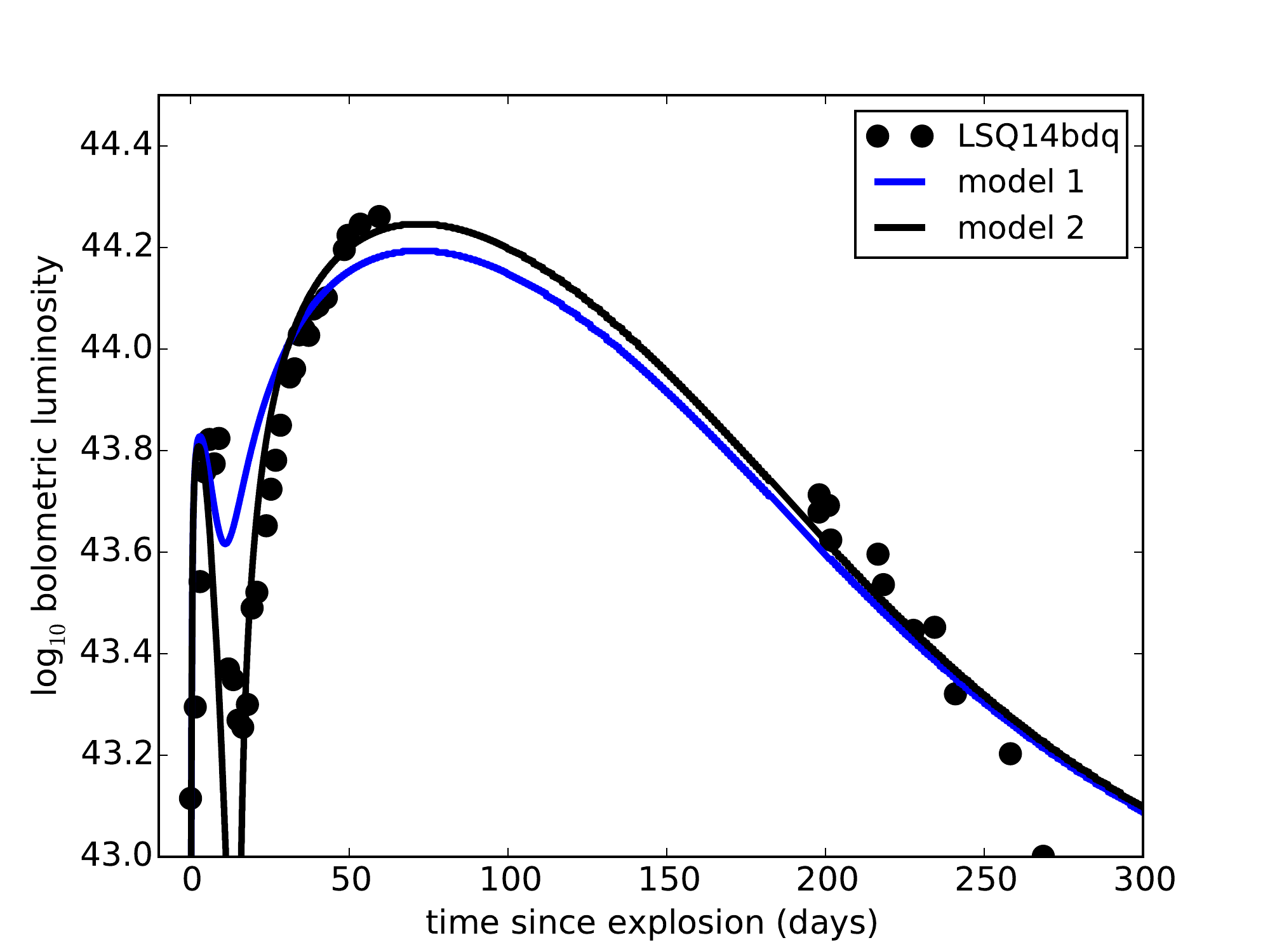}
\caption{
Comparison of the observed bolometric light curve of \SNd\ (black circles) to model light curves with parameters $\Msn = 25~M_\odot, \Esn = 5 \times 10^{51}~{\rm ergs}, \tm = 1.5$~days, and $\Em = 5 \times 10^{52}$~ergs.   Model 1 (blue lines) assumes that the magnetar wind energy is inefficiently thermalized as described by Equation~\ref{eq:Lth}. Model 2 (black lines) simply assumes that the thermalization is completely inefficient for $t < 15$~days, and 100\% efficient thereafter. The later approach provides more contrast in the peaks.
   \label{fig:LC_fit} }
\end{figure}

The predicted luminosity and timescale of magnetar driven shock breakout are compellingly similar to those seen in the double-peaked light curves of some SLSNe.  As
a concrete example,  Figure~\ref{fig:LC_fit} compares  approximate light curve models to the observed bolometric light curve of \SNd\ (as constructed by \cite{Nicholl_2015}).  The models have $\Msn = 25~M_\odot, \Esn = 5 \times 10^{51}~{\rm ergs}, \tm = 1.5$~days, and $\Em = 5 \times 10^{52}$~ergs, although we do not claim these values to be optimal or unique.  The adopted magnetar energy is high,  but well within the possible range recently found for rapidly rotating neutron stars \citep{Metzger_2015}.  

To see a clear double peak in the models of Fig~\ref{fig:LC_fit} required that we assumed inefficient
magnetar heating at early times.  We tried two ways of implementing this: the first used the  suppression function  Eq.~\ref{eq:Lth}; the second simply assumed that heating was completely inefficient until a time $t = 15$~day, and 100\% efficient thereafter.  The latter results in a clearer separation of the two light curve peaks, in better agreement with the observations.  Clearly the uncertain details of magnetar wind thermalization are  important in 
setting the precise shape of the early time light curve. The failure of the model to fit the
observations at $t > 250$~days may also be due to a decrease in thermalization efficiency at late times (for distinct physical reasons).

The reasonable model fits shown in Figure~\ref{fig:LC_fit} suggests that the breakout
scenario holds
 some promise for explaining double-peaked SLSNe.  However one should not put too much weight on this comparison (or the inferred physical parameters) given the number of approximations that have gone into our model light curves.
In particular, we have made coarse assumptions regarding spherical symmetry, the efficiency of thermalization, and the treatment of the radiative transfer.   Detailed radiation hydrodynamical calculations are needed to make a more meaningful comparisons to data.

 If, in contrast to the models of Figure~\ref{fig:LC_fit}, thermalization is
efficient at early times,  the shock breakout peak is typically not easily distinguished from the diffusive luminosity from central magnetar heating.
To explain a double peaked light curve in this case requires a specific
sets of parameters -- a supernova with 
 a  large ejecta mass ($\Msn \gtrsim 15~M_\odot$) and/or a high explosion energy ($\Esn \gtrsim 3 \times 10^{51}$~ergs).
  For  high ejecta masses, one might expect  the collapsing stellar core  to form a black hole rather than a neutron star.  Most of our  discussion of driven shock emission applies equally  
to an  engine  powered by black hole accretion \citep{Dexter_Kasen_2013}, in which case the energy input is set by the rate of fallback.

Deviations from our assumption of spherical symmetry presumably affect the luminosity and
timescale of breakout.
The actual geometry is likely bipolar, as even a sub-equipartition
toroidal magnetic field can act through hoop stress to confine the nebular
pressure along the rotational axis.  The resulting anisotropic stress may
drive a weak, wide-angle ``jet"
(\citealt{Bucciantini+07}) and the shocked ejecta will  take the form of a broad ``cocoon" that enshrouds that jet (as has been discussed for normal gamma-ray bursts (GRBs), e.g., \citealt{Lazzati&Begelman05}).   
In SLSNe, the shock breakout  of this ``cocoon" would
emerge continuously over a timescale of several days or longer, as set by
the  engine duration and the large  size of the remnant.
 Our spherical analysis here
may still be used to roughly estimate the  dynamics, but with the input magnetar power  enhanced by a factor of $4 \pi/\Omega$ where $\Omega$ is the solid angle
of the ``jet".   

Jet-like collimation of the energy should presumably result in a brighter shock breakout,  at least for some viewing angles.  This may make double-peaked light curves conspicuous even when inefficient magnetar heating is not invoked.
The light curve will depend on orientation, with the breakout  emission being more prominent for polar viewing angles, and less so for equatorial views.  Because the jet cocoon is expected to be fairly broad, bipolar, and non-relativistic, at least some breakout emission is likely to be emitted in all directions.

We have not considered how energy is thermalized behind the magnetar driven shock.
 \citet{Katz_2010} show that, in supernova shock breakout,  the post-shock gas and radiation reach  equilibrium for shock velocities $v/c \lesssim 0.1$.  The 
velocities expected here are much lower, $\vs - \ve \approx 0.01c$. 
  Once the shock emerges, however,  and  if it continues to be driven into low density, optically thin ejecta, equilibrium may no longer be reached.  In this case, one could look for some fraction of the shock heating emerging as non-thermal x-ray or radio emission.  Figure~\ref{fig:shock} shows that, for typical parameters, the shock heating rates are $\sim 10^{43}-10^{44}~\ergss$ at day~10, dropping to
  $10^{41}-10^{42}~\ergss$ by days 50-100.

The dynamical effect of  magnetar energy injection has additional observational consequences.  
An abrupt  but small increase in the Doppler shifts of line absorption features (by an amount $\approx 1000~\kms$)
may occur when the photosphere recedes into the region of non-monotonic ejecta velocities created by the shock (see Figure~\ref{fig:hydro}).
Following this, the photospheric velocity should coincide with the motion of the swept up shell.     For  long spindown times, the shell may still be accelerating at the time spectra are taken, such that the photospheric velocity  {\it increases} with time, counter to the
behavior expected for free expansion. Such an effect may have been observed in 
the helium lines of SN2005bf \citep{Tominaga_2005}, a double-peaked Type~Ib supernova that \cite{Maeda_2007} modeled with magnetar heating.    Once the magnetar has shut off,
the shell and photospheric velocity should approach a constant value over time.
These expectations  may have to be modified 
to account for asymmetries due to an anisotropic  magnetar wind
or hydrodynamical instabilities in the shell.

The magnetar model is but one explanation of double-peaked SLSN light curves.  While the first peak  in \SNd\  was too brief and bright to be explained by \Nifs\ heating, possible alternative mechanisms include cooling emission from a hyper-energetic SN explosion  \citep{Nicholl_2015} or  interaction with a dense CSM \citep{Moriya_2012,Piro_2015}. 
In either case the necessary ejecta kinetic energy is  large, $4-50 \times 10^{51}$~ergs, depending on
the assumed radius of the star or CSM shell \citep{Nicholl_2015, Piro_2015}.  
The second light curve peak  requires a distinct mechanism, 
either central engine heating or interaction with an additional CSM shell at larger
radius. While such a multi-component scenario can not be ruled out, 
it is appealing that the magnetar model alone may be able to  reproduce the double-peak without introducing additional model parameters, or requiring
extreme  values of the existing ones.    

Determining the fraction of SLSNe with double-peaked light curves would help
discriminate the mechanism responsible.  Models that explain both peaks with CSM interaction require two distinct CSM shells -- one low mass, nearby shell and one higher mass, more distant shell.  There is no obvious reason why pre-supernova mass loss would  conspire to frequently produce such a configuration. In the magnetar model, on the other hand,  some early time emission from shock breakout is a generic consequence of the central energy injection.
  
Spectra taken at the time of the first peak would also be diagnostic.
 Magnetar driven shock breakout  is expected to produce 
a blue ($T_{\rm eff} \approx 20,000$~K) quasi-blackbody spectrum that is  mostly featureless due to the high temperature and ionization state.  Any detectable line features would be of high velocity ($\gtrsim 10,000~\kms$).  In the CSM interaction models, in contrast, one might expect narrow line emission from a photo-ionized, slowly moving CSM shell, or perhaps  narrow line absorption if a second, cold CSM shell exists at larger radius.

While the magnetar model has been most frequently invoked to explain SLSNe,   a lower level of magnetar powering may  occur in less luminous supernovae, perhaps in some cases  producing a double peaked light curve.
For example, for  supernovae where the  explosion energy and magnetar energy are both $ \approx 10^{50}$~ergs, the predicted shock breakout
luminosity is only  $\approx  10^{42}$~\ergss, but produces a noticeable early peak in the light curve.

If double-peaked SLSNe light curves are indeed due to shock breakout, this may indicate an interesting
connection with the recently discovered class of very long duration  GRBs.  
\cite{Greiner_2015} present observations of a $\sim 10^4$~s long GRB which had an associated super-luminous optical transient, both of which they argue are powered by a magnetar. \cite{Metzger_2015} suggested that the jet in this event was just barely powerful enough to escape the stellar remnant.  For events with longer magnetar spindown timescales  $\gtrsim 10^4$~s, or  higher ejecta masses, a relativistic GRB jet may fail to emerge, however the underlying engine may still be revealed by the breakout of the magnetar driven shock (or ``cocoon") producing a double-peak optical light curve.

While the calculations in this paper have outlined the main features of magnetar driven shock breakout,  radiation-hydrodynamical calculations that include realistic opacities and, ideally, magnetic fields and departures from spherical symmetry, are needed for detailed predictions.
The observational constraints on the magnetar model are now many: the timescale and  luminosity of the first peak, the shape
and brightness of the second peak,
the photospheric velocity evolution, and the luminosity, color,  and decline rate of the  late time tail emission.
Simultaneous fitting of all of these observables within the limited model parameters $(B, P, \Msn, \Esn)$ constitutes
a non-trivial test of the  paradigm. 
\\  
\acknowledgements We thank Roger Chevalier, Tony Piro,  Eliot Quataert, Stephen Smartt, Tuguldor Sukhbold, and Stan Woosley for helpful conversations and comments on the draft, and Matt Nicholl for providing the bolometric light curve data for \SNd.
DK is supported in part by a Department of Energy Office of Nuclear
Physics Early Career Award, and by the Director, Office of Energy
Research, Office of High Energy and Nuclear Physics, Divisions of
Nuclear Physics, of the U.S. Department of Energy under Contract No.
DE-AC02-05CH11231. 
BDM gratefully acknowledges support from the NSF grant AST-1410950 and the Alfred P. Sloan Foundation.
  This work was supported by
the National Science Foundation under grants PHY 11-25915,
AST 11-09174, and AST 12-05574.
This work was supported in part by NSF Grant No. PHYS-1066293 and the hospitality of the Aspen Center for Physics.
\appendix

To calculate approximate light curves, we  use a one-zone formalism \citep{Arnett_1982} which has frequently been applied to model magnetar powered light curves \citep[e.g.,][]{Kasen_2010, Inserra_2013}.  
To model the supernova light curve powered by direct magnetar heating we 
consider the evolution of the  internal energy, $E_{\rm int}$, of the bulk of the ejecta 
\begin{equation}
\frac{d E_{\rm int}}{d t} = -p \frac{d V}{d t} + \Lm - \Lsn
\label{eq:onezone}
\end{equation}
where $V$ is the volume, $p$ is the pressure.
In the diffusion approximation, the radiated luminosity is
\begin{equation}
	\Lsn = 4 \pi R^2 \frac{c}{3 \kappa \rho} \frac{\partial (E_{\rm int}/V)}{\partial R} 
	\approx \frac{4 \pi c R^2}{3 \kappa \rho} \frac{(E_{\rm int}/V)}{R}
\end{equation}
Assuming homologous expansion ($R = vt$) gives
\begin{equation}
\Lsn \approx E_{\rm int} t/\tsn^2~~{\rm where}~\tsn  = \left[ \frac{ 3 \kappa \Msn}{ 4 \pi v c} \right]^{1/2}
\end{equation}
The timescale \tsn\ is similar to the diffusion time \td\ (Eq.~\ref{eq:td}) although here we include 
the energy input by the magnetar to calculate the velocity, $v = [ 2 (\Esn + \Em)/\Msn]^{1/2}$. 

The formal solution to the differential equation Eq.~\ref{eq:onezone} is 
\begin{equation}
\Lsn = e^{-(t/\tsn)^2} \int_0^t 2 \Lm (t'/\tsn)  e^{-(t'/\tsn)^2} dt',
\label{eq:solution}
\end{equation}
The treatment here is clearly  approximate, as it neglects the  formation and expansion of the shell structure.
Previous calculations have shown, however, that the one-zone formalism well reproduces the
light curves from more detailed radiation-hydrodynamical models \citep{Inserra_2013}.

 To calculate the luminosity due to magnetar driven shock heating, we solve an independent one zone model, using the same integral expression Eq.~\ref{eq:solution}
but with the heating rate
\Lm\ replaced with the shock heating rate \epss\ (from Eq.~\ref{eq:esh}).  The time dependent \epss\ was determined from our numerical
integration of the shell dynamics (see Figure~\ref{fig:shock}).
We further replace  \tsn\  with  \tbo\ (Eq.~\ref{eq:tbo}), as \tbo\ gives the appropriate timescale when the diffusion time from
the shock heated region equals the elapsed time.  In these calculations we use Eq.~A\ref{eq:solution} to include the radiative loss term in the
shell evolution, i.e. $\Lsn \ne 0$  in the energy equation (Eq.~\ref{eq:dynE}).   The two one-zone model light curves were summed to give the composite supernova light curve.

\bibliographystyle{apj}
\bibliography{ms}

 \end{document}